\documentstyle[pre,aps]{revtex}

\begin{document}

\draft

\title{Nonlinear Dynamics of Nuclear--Electronic Spin Processes in
Ferromagnets} 

\author{V.I. Yukalov$^{1,2}$, M.G. Cottam$^1$, and M.R. Singh$^1$} 

\address{$^1$ Center for Chemical Physics \\
University of Western Ontario, London, Ontario N6A 3K7, Canada \\ [2mm]
$^2$ Bogolubov Laboratory of Theoretical Physics \\
Joint Institute for Nuclear Research, Dubna 141980, Russia}

\maketitle

\begin{abstract}

Spin dynamics is considered in ferromagnets consisting of electron and
nuclear subsystems interacting with each other through hyperfine forces.
In addition, the ferromagnetic sample is coupled with a resonance electric
circuit. Under these conditions, spin relaxation from a strongly nonequilibrium
initial state displays several peculiarities absent for the standard set--up 
in studying spin relaxation. The main feature of the nonlinear spin dynamics
considered in this communication is the appearance of ultrafast coherent
relaxation, with characteristic relaxation times several orders shorter
than the transverse relaxation time $T_2$. This type of coherent spin
relaxation can be used for extracting additional information on the
intrinsic properties of ferromagnetic  materials and also can be employed
for different technical applications.

\end{abstract}

\newpage

{\parindent=0pt

{\bf I. INTRODUCTION}}

\vspace{3mm}

Spin systems can exhibit rather nontrivial dynamics when the magnetic
sample is prepared in a strongly nonequilibrium state and, in addition, is
coupled with a resonance electric circuit [1]. Due to the resonator
feedback field, a coherent motion of spins can develop resulting in their
ultrafast relaxation. However, the feedback field can organize coherent
relaxation only when some initial mechanism triggers the process. A simple
case could be the application of an external pulse at the initial time. If
this pulse is sufficiently strong, the spin dynamics could be described by
the Bloch equations. A more difficult, but interesting, situation is when
no external pulse starts the process, but the latter develops in a
self--organized way due to local spin fluctuations caused by their
interactions. In such a case, the Bloch equations are not appropriate [1]
and one has to resort to microscopic models.

A theory of coherent spin relaxation in a system of nuclear spins inside a
paramagnetic matrix has been developed [2,3], being based on a microscopic
Hamiltonian with dipole interactions between nuclei. In the present paper
we generalize this theory to include ferromagnetic materials. As far as
ferrimagnets are often described as ferromagnets with an effective
magnetization, our approach is applicable to ferrimagnets as well. In this
way, our aim is to suggest a general microscopic theory valid for a wide
class of materials, including those having long--range magnetic order.

\vspace{5mm}

{\parindent=0pt

{\bf II. THEORY}}

\vspace{3mm}

We consider a magnet consisting of electronic and nuclear
subsystems interacting with each other by hyperfine forces. The system of
electrons possesses long--range ferromagnetic order. The subsystem of
nuclear spins is prepared in a strongly nonequilibrium state, which can be
done by dynamic nuclear polarization techniques. The sample is inserted
into a coil connected with a resonance electric circuit. The general
Hamiltonian describing a wide class of magnetic materials can be
taken in the form
\begin{equation}
\hat H =\hat H_e + \hat H_n + \hat H_{int} \; ,
\end{equation}
in which
\begin{equation}
\hat H_e = -\frac{1}{2}\sum_{i\neq j} J_{ij}{\bf S}_i
\cdot {\bf S}_j -\mu_e\sum_i {\bf B} \cdot {\bf S}_i
\end{equation}
is the Hamiltonian of electron spins,
\begin{equation}
\hat H_n = \frac{1}{2}\sum_{i\neq j}
\sum_{\alpha\beta} C_{ij}^{\alpha\beta} I_i^\alpha I_j^\beta - 
\mu_n\sum_i{\bf B} \cdot{\bf I}_i
\end{equation}
is the nuclear spin Hamiltonian, and
\begin{equation}
\hat H_{int} = A \sum_i{\bf S}_i\cdot
{\bf I}_i + \frac{1}{2}\sum_{i\neq j}
\sum_{\alpha\beta} A_{ij}^{\alpha\beta} S_i^\alpha I_j^\beta
\end{equation}
is the term corresponding to hyperfine interactions. Here $J_{ij}$ is an
exchange interaction; $\mu_e =g_e\mu_B$, with $g_e$ being the electron
gyromagnetic ratio and $\mu_B$, the Bohr magneton; the nuclear dipole
interactions $C_{ij}^{\alpha\beta} = \mu_n^2 \left ( \delta_{\alpha\beta} - 
3n_{ij}^\alpha n_{ij}^\beta\right )/r^3_{ij}$
contain $\mu_n=g_n\mu_N$, with $g_n$ being the nuclear gyromagnetic ratio
and $\mu_N$, nuclear magneton, and $r_{ij}\equiv |{\bf r}_{ij}|,\; 
{\bf n}_{ij}\equiv {\bf r}_{ij}/r_{ij},\; {\bf r}_{ij}\equiv {\bf r}_i-
{\bf r}_j$; the hyperfine interactions consist of a contact part with a
constant $A$ and of a dipole-dipole part with
$A_{ij}^{\alpha\beta} = \mu_e\mu_n \left ( \delta_{\alpha\beta} -
3n_{ij}^\alpha n_{ij}^\beta\right )/r^3_{ij}$;
the indices $i$ and $j$ enumerate electrons or nuclei according to the
context, and $\alpha,\beta=x,y,z$; ${\bf S}_i$ is an electron spin
operator while ${\bf I}_j$ is a nuclear spin operator. The total magnetic
field ${\bf B}$
is the vector sum of an external magnetic field in the $z$ direction and of a
transverse field $H_1=H_a+H$ in the $x$ direction, consisting of an effective field $H_a$ of a
transverse magnetocrystalline anisotropy and of a resonator feedback field
$H$. The latter satisfies the Kirchhoff equation
\begin{equation}
\frac{dH}{dt} + 2\gamma_3 H +\omega^2 \int_0^t H(\tau)d\tau = 
-4\pi\eta\frac{dM_x}{dt}\; ,
\end{equation}
in which $\eta$ is a filling factor; $\omega$, resonator natural frequency;
$\gamma_3\equiv\omega/2Q$ is the resonator ringing width; $Q$ is a quality
factor; and $M_x=\frac{1}{V}\sum_i(\mu_e < S_i^x >+\mu_n<I_i^x>)$ is the 
transverse magnetization, where the angle brackets mean the statistical
averaging.

Employing the Heisenberg equations of motion, we derive the time evolution
equations for the following averages, related to the electron and nuclear
spins,
$$
x\equiv \frac{1}{N_e} \sum_i < S_i^- > \; , \qquad
z\equiv \frac{1}{N_e} \sum_i < S_i^z > \; ,
$$
\begin{equation}
u\equiv \frac{1}{N_n} \sum_i < I_i^- > \; , \qquad
s\equiv \frac{1}{N_n}\sum_i < I_i^z > \; ,
\end{equation}
where $N_e$ and $N_n$ are the number of electrons and nuclei, respectively, 
and $S_i^-$ and $I_i^-$ are the ladder operators.
As the transverse variables $x$ and $u$ are complex, we need
also the equations of motion for either $x^*$ and $u^*$ or $|x|$ and
$|u|$. In this way, we obtain seven evolution equations, three for the
electron variables $x,\; z$ and $|x|^2$, three for the nuclear variables
$u,\; s$, and $|u|^2$, and the feedback--field equation (5). Although this
is a rather complicated system of nonlinear equations, it can be treated
by using the scale separation approach [2,3]. Details of this approach
have been thoroughly described in Refs. [3--5].

Note first that if one invokes the standard semiclassical decoupling of
spin correlators, assuming the translational invariance of the average
spins, then some of the terms in the evolution equations become zero
because of the properties of the dipolar interactions. The translation
invariance of averages is equivalent to neglecting inhomogeneous local spin 
fluctuations. However, the latter are crucially important for the correct
description of spin dynamics [1--3]. The inhomogeneous spin fluctuations
can be retained by treating them as random local fields. Thus we come to
the stochastic semiclassical approximation [3--5]. Then, using the method
of the Laplace transforms, we may express the feedback field from Eq. (5)
through the derivatives of spin variables and employ this relation in the
evolution equations for $x,\; z,\; |x|^2$, and $u,\; s,\; |u|^2$. The
latter spin variables can be classified into fast and slow with respect to
each other by comparing their time derivatives. We keep in mind the
following usual inequalities:
\begin{equation}
\left | \gamma_1{\Large /} \omega_E \right | \ll 1\; , \qquad
\left | \gamma_2{\Large /} \omega_E \right | \ll 1\; , \qquad
\left | \Gamma_1{\Large /} \omega_N \right | \ll 1\; ,\qquad
\left | \Gamma_2{\Large /} \omega_N \right | \ll 1 \; ,
\end{equation}
in which $\gamma_1$ and $\Gamma_1$ are the transverse attenuations for the
electron and nuclear spins, $\gamma_2$ and $\Gamma_2$ are the longitudinal
attenuations for electron and nuclear spins, respectively, and
\begin{equation}
\omega_E \equiv \left ( \mu_e H_0 - As\right ){\Large /}\hbar \; , \qquad
\omega_N \equiv \left ( \mu_n H_0 - Am\right ){\Large /}\hbar \; 
\end{equation}
are the electron spin resonance frequency and the nuclear magnetic
resonance frequency, respectively; $m$ being an average magnetization in
the electron system. Another reasonable assumption is that the external
magnetic field $H_0$ is stronger than the magnetocrystalline anisotropy
field $H_a$ and that, similarly to (7), the inhomogeneous broadening,
caused by local spin fluctuations, is smaller than the corresponding
frequencies, so that
\begin{equation}
\left | \alpha_e{\Large /}\omega_E \right | \ll 1\; , \qquad
\left | \alpha_n{\Large /}\omega_N \right | \ll 1\; , \qquad
\left | \gamma_*{\Large /}\omega_E \right | \ll 1\; , \qquad
\left | \Gamma_*{\Large /}\omega_N \right | \ll 1\; ,
\end{equation}
where $\alpha_e\equiv\mu_e H_a/\hbar$ and $\alpha_n\equiv\mu_n H_a/\hbar$
are the anisotropy parameters and $\gamma_*$ and $\Gamma_*$ are the
inhomogeneous widths for electrons and nuclei, respectively. Also, because
the nuclear magneton is three orders smaller than the Bohr magneton, we
have
\begin{equation}
\left | \mu_n{\Large /}\mu_e \right | \ll 1\; , \qquad
\Gamma_1{\Large /}\gamma_1  \ll 1\; , \qquad  
\Gamma_2{\Large /}\gamma_2 \ll 1\; .
\end{equation}
Finally, we consider the case of a high quality resonator, having a large
quality factor, and we assume that the resonator natural frequency is
tuned close to the frequency of nuclear magnetic resonance, so that
$|\Delta_N/\omega_N|\ll 1$ and $\gamma_3/\omega\ll 1$, where $\Delta_N=
\omega-\omega_N$.

With these inequalities, we can classify all variables into fast and slow
with respect to each other. Following the general scheme [3--5], we
solve the equations for fast variables while treating slow variables as
quasi--integrals of motion. Then, the found solutions for fast variables
are substituted into the equations for slow variables and the right--hand
sides of the latter equations are averaged over the periods of fast
oscillations and over the random local fields. Introducing also the change
of variables
\begin{equation}
w = |u|^2  - \frac{\alpha_n^2 +\Gamma_*^2 +\delta^2}{\omega_N^2} s^2 \; ,
\qquad \delta\equiv
\frac{\sqrt{2}\pi^2\eta\gamma_*\rho_e\mu_e\mu_n}{\omega_N} m \; ,
\end{equation}
we come to the equations describing the slow nuclear spin variables
\begin{equation}
\frac{ds}{dt} = \Gamma_2 gw - \Gamma_1 (s -\zeta) \; , \qquad
\frac{dw}{dt} = - 2\Gamma_2 ( 1 + gs) w \; ,
\end{equation}
where
\begin{equation}
g\equiv \pi^2\eta\frac{\rho_n\mu_n^2\omega_N}{\Gamma_2\omega} \left ( 1 +
\frac{\rho_e\mu_e Am}{\rho_n\mu_n\omega_N}\right )
\end{equation}
is the parameter of effective coupling of nuclear spins with the
resonator.

For the relaxation times $T_1\equiv\Gamma_1^{-1}$ and
$T_2\equiv\Gamma_2^{-1}$ one usually has the relation $T_2\ll T_1$.
Therefore, for the times $t\ll T_1$, equations (12) can be solved
analytically giving
\begin{equation}
s =\frac{T_2}{g\tau_0}{\rm tanh}\left (\frac{t-t_0}{\tau_0} \right ) -
\frac{1}{g} \; , \qquad
w =\left (\frac{T_2}{g\tau_0}\right )^2{\rm sech}^2\left (
\frac{t-t_0}{\tau_0} \right ) \; ,
\end{equation}
where $\tau_0$ is the collective relaxation time and $t_0$ is the delay time,
respectively,
\begin{equation}
\tau_0 =\frac{T_2}{\sqrt{(1+gs_0)^2 +g^2w_0^2}} \; , \qquad
t_0 = \frac{\tau_0}{2}\ln\left |
\frac{T_2-\tau_0(1+gs_0)}{T_2+\tau_0(1+gs_0)}\right | \; ,
\end{equation}
with $s_0=s(0)$ and $w_0=w(0)$ defined by initial conditions.

\vspace{5mm}

{\parindent=0pt

{\bf III. CONCLUSION}}

\vspace{3mm}

We analysed the obtained solutions for the parameters typical of such
ferromagnetic materials as EuO, EuS, EuSe, Li$_x$Fe$_{3-x}$O$_4$,
Mn$_x$Sb$_{1-x}$, NiMnSb, NiMnSi, Co$_2$MnSi, and Co in the fcc
and hcp phases [6,7]. Since ferrimagnets can often be modeled as
ferromagnets with an effective magnetization [8], our analysis is
applicable as well to ferrimagnetic materials, such as MnFe$_2$O$_4$. For
these materials, taking as initial conditions $s_0=-I,\; w_0=0$, where $I$
is a nuclear spin, we obtain the relaxation time $\tau_0 \cong T_2/gI$,
where $g\cong\pi^2\mu_e/2\mu_n$, and the delay time 
$t_0=\tau_0\ln(2\omega_N/10^7)$. This gives the coupling parameter 
$g\sim 10^4$, which, with $T_2\sim 10^{-4}$ s, yields the relaxation time 
$\tau_0\sim 10^{-8}$ s. Because $\omega_N\sim10^9$ s$^{-1}$, we have the 
delay time $t_0\sim 5\times 10^{-8}$ s.

The described regime corresponds to the ultrafast coherent spin relaxation, 
when during the time $t_0+\tau_0$ the initial strongly nonequilibrium spin
polarization $s_0=-I$ changes to its equilibrium value $s(t)\cong I$ at
$t\gg t_0$. The time $t_0+\tau_0$ can be four orders less than the
standard spin--spin relaxation time $T_2$. This ultrafast coherent spin
relaxation can serve as an additional technique for studying the spin-spin 
correlations in ferromagnetic materials, complementing other known
techniques, such as neutron diffraction, light scattering, and
nuclear magnetic resonance. The ultrafast relaxation mechanism can
also find application in the important problem of fast repolarization of
solid--state targets used in scattering experiments [9,10], as well as
for fast switching devices in electronics and computing.

\vspace{5mm}

{\bf ACNOWLEDGEMENT}

\vspace{3mm}

Financial support from the University of Western Ontario and NSERC of
Canada is appreciated.

\end{document}